%% file: eval_flaws.tex
\documentclass[manuscript]{acmart}
\usepackage{multirow}
\usepackage{hyperref}
\usepackage{graphicx}
\usepackage{subcaption}
\usepackage{hyperref}

\AtBeginDocument{%
  }

\copyrightyear{2023}
\acmYear{2023}
\setcopyright{acmlicensed}\acmConference[RecSys '23]{Seventeenth ACM Conference on Recommender Systems}{September 18--22, 2023}{Singapore, Singapore}
\acmBooktitle{Seventeenth ACM Conference on Recommender Systems (RecSys '23), September 18--22, 2023, Singapore, Singapore}
\acmPrice{15.00}
\acmDOI{10.1145/3604915.3608839}
\acmISBN{979-8-4007-0241-9/23/09}

\begin{document}

\title{Widespread Flaws in Offline Evaluation of Recommender Systems}

\author{Bal\'{a}zs Hidasi}
\email{balazs.h@taboola.com}
\author{\'{A}d\'{a}m Tibor Czapp}
\email{adam-tibor.c@taboola.com}
\affiliation{%
  \institution{Gravity R\&D, a Taboola company}
  \streetaddress{Vill\'{a}nyi \'{u}t 40/B}
  \city{Budapest}
  \country{Hungary}
  \postcode{1113}
}

\renewcommand{\shortauthors}{Hidasi and Czapp}

\begin{abstract}
  \input{0_abstract}
\end{abstract}




\maketitle

\input{1_intro}

\input{2_eval_flaws}
\input{3_related_work}
\input{4_conclusion}

\begin{acks}
  The work leading to these results received funding from \grantsponsor{PIACIKFI}{National Research, Development and Innovation Office, Hungary}~~ under grant agreement number \grantnum{PIACIKFI}{2020-1.1.2-PIACI-KFI-2021-00289}.

  The authors would also like to thank Domonkos Tikk for his valuable support as the project leader of the above-mentioned R\&D grant.
\end{acks}

\bibliographystyle{ACM-Reference-Format}
\bibliography{references}

\end{document}

%% file: 0_abstract.tex
Even though offline evaluation is just an imperfect proxy of online performance -- due to the interactive nature of recommenders -- it will probably remain the primary way of evaluation in recommender systems research for the foreseeable future, since the proprietary nature of production recommenders prevents independent validation of A/B test setups and verification of online results. Therefore, it is imperative that offline evaluation setups are as realistic and as flawless as they can be. Unfortunately, evaluation flaws are quite common in recommender systems research nowadays, due to later works copying flawed evaluation setups from their predecessors without questioning their validity. In the hope of improving the quality of offline evaluation of recommender systems, we discuss four of these widespread flaws and why researchers should avoid them.

%% file: 1_intro.tex
\section{Introduction}\label{sec:intro}

Evaluation is one of the biggest challenges of recommender systems research. Online A/B testing is already an approximation of the recommender's true goal, limited by KPIs that can be measured. Splitting traffic fairly, preventing information leak between groups and eliminating biases~\cite{jeunen2023common} are also non-trivial. From the business perspective, A/B tests are expensive and slow because a portion of the traffic is served by suboptimal models, and getting statistically significant differences for informative but noisy KPIs can take weeks/months. From the scientific perspective, A/B tests are not reproducible due to the proprietary nature of production recommenders preventing independent validation of A/B test setups and verification of online results. Therefore, the inherently imperfect approximation of online performance through offline evaluation is likely to stay the main way of assessing performance in research, at least until simulators~\cite{mcinerney2021accordion} become more mature. Thus, it is imperative that offline evaluation is correct and follows how production recommenders work as closely as possible. Unfortunately, flaws are quite common, due to later works copying flawed evaluation setups from their predecessors without questioning their validity~\cite{ferrari2020methodological}.

In this paper we discuss four widespread evaluation flaws we observed en masse in research papers of the last decade. Our main contribution is presented in Section~\ref{sec:flaws}, where we go through the most important steps of designing offline evaluation setups and discuss common pitfalls. We closely examine four severe flaws commonly found in the evaluation setup of various recommenders: dataset--task mismatch, general claims on heavily preprocessed data, information leaking through time, and negative item sampling. Their effect is demonstrated on sequential recommenders.

%% file: 2_eval_flaws.tex
\section{Design flaws in evaluation setups}\label{sec:flaws}

In this section we go through the main steps of designing offline evaluation and point out potential pitfalls and flaws.

Recommender systems are utilized in various scenarios covering a wide spectrum of domains, use-cases, data and goals, which might require completely different approaches. Therefore, step zero is task definition, since the offline evaluation is designed around the task, not vice versa. The actual first design step is deciding on the evaluation methodology and metrics. Selecting options that are inappropriate for the task undermines the whole evaluation. 

The two main methodologies differ in how they model user preference. Under \emph{behavior prediction}, user preferences are described by both interactions with the recommender and organic events. Model performance is measured by how well it can predict user behavior. The research community has been utilizing this setup since the early days of implicit feedback based recommenders (e.g.~\cite{hu2008collaborative}). This methodology suffers from the lack of negative feedback. It is also imperfect since flawless behavior prediction would recommend the same items that the user would have interacted with on their own. However, it is still a satisfactory setup because the behavior modeling capability of an algorithm loosely correlates with online performance. \emph{Interaction prediction} only considers the interactions with the recommender. Positive/negative feedback is generated when a user does/doesn't interact with a recommended item. The traditional use-case of this methodology has been CTR prediction~\cite{guo2017deepfm}. The main challenge is that reliance on bandit feedback results in a closed feedback loop, making it hard to estimate the uplift of a new algorithm through off-policy evaluation. Counterfactual evaluation~\cite{saito2021counterfactual} aims at alleviating the problem, but still does not scale up to the large state--action space of recommenders. 

The two main options for metrics are ranking/IR (e.g.~recall@N, MRR@N, NDCG@N, etc.) and classification metrics (e.g.~AUC, accuracy, etc.). Behavior prediction often uses the former and interaction prediction the latter, but this is not a necessity. Auxiliary metrics (e.g.~diversity, novelty, serendipity) can be used along with accuracy metrics, and recent research shows that serendipity can have a significant impact on online performance~\cite{chen2021values}. 

Common flaws of subsequent design steps are discussed in detail. These flaws are generic, but for demonstration purposes, we chose the next item prediction of session-based recommenders~\cite{hidasi2015session} as the task, behavior prediction as the appropriate methodology, and recall@N and MRR@N as offline metrics. Most of our experiments utilize the official implementation\footnote{\url{https://github.com/hidasib/GRU4Rec}} of GRU4Rec\cite{hidasi2015session,hidasi2018recurrent}, because the speed-optimized code allows for quick experimentation on both small and large datasets. The code of our experiments, hyperparameters and additional results are publicly available\footnote{\url{https://github.com/hidasib/recsys_eval_flaws}}.

\subsection{Dataset--task mismatch}\label{ssec:flaw-data}

The next step is choosing datasets. For the sake of reproducibility, public datasets should be used along with or instead of proprietary ones. This is where a common flaw can happen. Not every dataset is appropriate for every task or evaluation methodology. Data might be transformed to somewhat accommodate a task, e.g.~rating data was often used as implicit feedback (e.g.~\cite{pilaszy2010fast,hidasi2012fast}), when public implicit feedback datasets were not available. However, the validity of this transformation can be questionable. If all ratings are treated as implicit feedback, negative feedback is knowingly treated as positive; on the other hand, if only high ratings are treated as implicit feedback, it becomes artificially less noisy than feedback in real-life datasets. Nowadays more and more datasets~\cite{requena2020shopper,kang2018self,wan2018item} are made available for a wide array of recommendation tasks. However, many researchers still use the same datasets regardless of the task.

Sequential recommendation -- i.e.~when the collections of events (e.g.~user histories, sessions) are treated as sequences -- only makes sense if the data has sequential patterns. E.g.~sessions of similar topics happening within a few days often have similar patterns because the service with which the users interact sets them on similar paths. \cite{petrov2022systematic} summarized that the most commonly (e.g.~\cite{kang2018self,sun2019bert4rec,huang2018improving,zhou2020s3}) used datasets for evaluating sequential recommenders are rating datasets\footnote{ML10M: \url{https://grouplens.org/datasets/movielens/10m/}; Steam: \url{https://cseweb.ucsd.edu/~jmcauley/datasets.html\#steam\_data}; Yelp: \url{https://www.yelp.com/dataset}; Amazon: \url{http://jmcauley.ucsd.edu/data/amazon/links.html}} (MovieLens, Steam, Yelp, Amazon (Beauty)) wherein user rating histories are treated as sequences. However, the presence of sequential patterns is questionable, because the time of rating is disjoint from the time of interacting with the item. E.g.~the user might skip rating  items they have no strong opinion on.
  
\begin{table*}
  \caption{Basic statistics of train/test splits and event collision rate of the datasets}
  \label{tab:data_stat}
  \footnotesize
  \begin{tabular}{l|rrr|rrr|r||rr}
    \toprule
    \multirow{2}{*}{Dataset} & \multicolumn{3}{c|}{Training set} & \multicolumn{3}{c|}{Test set} & \multirow{2}{*}{\#Items} & \multicolumn{2}{c}{Event time collisions}\\
    & \multicolumn{1}{c}{\#Events} & \multicolumn{1}{c}{\#Sequences} & \multicolumn{1}{c|}{\#Days} & \multicolumn{1}{c}{\#Events} & \multicolumn{1}{c}{\#Sequences} & \multicolumn{1}{c|}{\#Days} & &  \multicolumn{1}{c}{Proportion} & \multicolumn{1}{c}{Event\%}\\
    \midrule
    Amazon (Beauty) & 724,440 & 215,595 & 4,907 & 30,191 & 11,452 & 56 & 38,606 & 31.89\% & 33.03\% \\
    MovieLens10M & 9,861,612 & 69,141 & 5,054 & 99,022 & 737 & 56 & 10,066 & 17.83\% & 27.33\% \\
    Steam & 4,856,479 & 900,878 & 2,582 & 46,039 & 16,916 & 56 & 12,229 & 7.67\% & 13.49\% \\
    Yelp & 5,583,947 & 810,015 & 6,091 & 15,437 & 5,183 & 91 & 132,895 & 0.05\% & 0.06\% \\
    \midrule
    Rees46 & 67,575,203 & 10,190,006 & 60 & 1,054,210 & 166,841 & 1 & 172,756 & 0.03\% & 0.04\% \\
    Coveo & 1,411,113 & 165,673 & 17 & 52,501 & 7,748 & 1 & 10,868 & 0.00\% & 0.00\% \\
    RetailRocket & 750,832 & 196,234 & 131 & 29,148 & 8,036 & 7 & 36,824 & 0.05\% & 0.05\% \\
    \bottomrule
  \end{tabular}
\end{table*}

The aforementioned datasets are compared to three session datasets\footnote{Rees46: \url{https://www.kaggle.com/datasets/mkechinov/ecommerce-behavior-data-from-multi-category-store}; Coveo: \url{https://github.com/coveooss/shopper-intent-prediction-nature-2020}; Retailrocket: \url{https://www.kaggle.com/datasets/retailrocket/ecommerce-dataset}} -- Rees46, Coveo~\cite{requena2020shopper} and RetailRocket -- to investigate the presence of sequential patterns. Data preprocessing is as follows (see Table~\ref{tab:data_stat} for basic statistics):
\begin{enumerate}
    \item Session datasets might contain multiple event types. If so, only the one corresponding to item views is kept.
    \item Sequences are corresponding to user histories in rating datasets, precomputed sessions in Coveo, and sessions with one hour session gap in Rees46 and RetailRocket.
    \item Subsequent repeating items are removed from the sequences, e.g.~$(i,i,j)\rightarrow(i,j)$, but $(i,j,i)$ is unchanged. These are not informative for recommenders because recommending the same item to the user is not useful.
    \item The dataset is iteratively filtered for sequences shorter than 2 and items occurring less than 5 times until there is no change to the dataset. Sequences consisting of one item are useless for this task. The weak item support filter is applied because of the collaborative filtering nature of the algorithms.
    \item Train/test splits are time based. Split time is set so that the size of the test data is satisfactory, but it is at least one day before the last event of the dataset. The test set consists of sequences that started after the split time. The train set consists of events that happened before the split time, i.e.~sequences extending over are cut off.
\end{enumerate}
Analyzing the data reveals that the resolution of timestamps is one day for the Amazon and Steam datasets, which might result in event collision, i.e.~two or more events of the same user having the same timestamp. The order of the events of an event collision can not be determined, and thus their sequence is unreliable and might be changed unintentionally during training or testing. The two rightmost columns in Table~\ref{tab:data_stat} show the proportion of collisions to all user--timestamp pairs and the proportion of participating events to all events. Beside Amazon and Steam, MovieLens10M also has many event collisions, which is extremely problematic, because this dataset comes presorted by user and item ID by default. This means that if user $A$ originally had a rating sequence on a set of items (e.g.~$k\rightarrow i \rightarrow j$), while user $B$ rated the same items in a different order (e.g.~$j\rightarrow k\rightarrow i$), then timestamp collision and the default sorting by item ID together produces the same sequence ($i\rightarrow j\rightarrow k$) for both. This is not the original sequence of user $A$ or $B$ but introduces two instances of an artificial sequence that did not even occur. On a larger scale, this phenomenon introduces artificial sequential patterns even if the data was not sequential originally.

\begin{table*}
    \caption{Recommendation accuracy using the same model with and without sequence modelling}
    \label{tab:seq_vs_noseq}
    \footnotesize
    \begin{tabular}{l|llll|llll|rrrr}
        \toprule
        Dataset & \multicolumn{4}{c|}{Model w/ sequence modelling} & \multicolumn{4}{c|}{Model w/o sequence modelling} &  \multicolumn{4}{c}{Relative change} \\
        & \multicolumn{2}{c}{Recall@N} & \multicolumn{2}{c|}{MRR@N} & \multicolumn{2}{c}{Recall@N} & \multicolumn{2}{c|}{MRR@N} & \multicolumn{2}{c}{Recall@N} & \multicolumn{2}{c}{MRR@N} \\
        & \multicolumn{1}{c}{N=5} & \multicolumn{1}{c}{N=20} & \multicolumn{1}{c}{N=5} & \multicolumn{1}{c|}{N=20} & \multicolumn{1}{c}{N=5} & \multicolumn{1}{c}{N=20} & \multicolumn{1}{c}{N=5} & \multicolumn{1}{c|}{N=20} & \multicolumn{1}{c}{N=5} & \multicolumn{1}{c}{N=20} & \multicolumn{1}{c}{N=5} & \multicolumn{1}{c}{N=20} \\    
        \midrule
        Rees46 & 0.3010 & 0.5293 & 0.1778 & 0.2008 & 0.2594 & 0.4785 & 0.1474 & 0.1694 & -13.80\% & -9.58\% & -17.09\% & -15.67\% \\
        Coveo & 0.1496 & 0.3135 & 0.0852 & 0.1010 & 0.1289 & 0.2678 & 0.0734 & 0.0868 & -13.83\% & -14.59\% & -13.85\% & -14.05\% \\
        Retailrocket & 0.3237 & 0.5186 & 0.1977 & 0.2175 & 0.2747 & 0.4652 & 0.1613 & 0.1806 & -15.13\% & -10.30\% & -18.42\% & -16.97\% \\
        Amazon (Beauty) & 0.0784 & 0.1319 & 0.0527 & 0.0579 & 0.0779 & 0.1271 & 0.0531 & 0.0579 & -0.71\% & -3.61\% & 0.86\% & 0.00\% \\
        MovieLens10M & 0.1728 & 0.3264 & 0.1062 & 0.1211 & 0.1276 & 0.2440 & 0.0763 & 0.0875 & -26.18\% & -25.23\% & -28.16\% & -27.68\% \\
        Steam & 0.1117 & 0.2371 & 0.0662 & 0.0781 & 0.1035 & 0.2208 & 0.0622 & 0.0735 & -7.38\% & -6.87\% & -5.99\% & -5.96\% \\    
        Yelp & 0.0702 & 0.1627 & 0.0371 & 0.0457 & 0.0657 & 0.1625 & 0.0353 & 0.0445 & -6.46\% & -0.12\% & -4.78\% & -2.51\% \\
        \bottomrule
    \end{tabular}
\end{table*}

Table~\ref{tab:seq_vs_noseq} shows recommendation accuracy of a sequential recommender (GRU4Rec) and the same algorithm with the sequence modeling part (GRU) replaced with a feedforward layer. Hyperparameters are optimized separately for the two algorithms and seven datasets, since replacing the GRU layer might alter the optimal parameters. A validation set  -- created from the full training set using the same process as the train/test split -- is used for optimization. Then the algorithm is retrained on the full training set using the optimal parameters, and performance is measured on the test set.

Results show that sequence modeling is not important for the Amazon and Yelp datasets\footnote{Small differences in offline metrics don't translate to any change in online performance due to the proxy nature of the offline setup.}, indicating the lack of sequential patterns and that they are unfit for evaluating sequential algorithms. Sequence modeling has a small positive impact on Steam, despite the daily resolution of the timestamp due to the specific user behavior of its domain. The rest of the datasets greatly benefit from session modeling, suggesting the presence of sequential patterns. However, in the case of MovieLens10M, these are artificial patters resulting from event collisions and presorting.

\subsection{Overzealous preprocessing}\label{ssec:flaw-preproc}

Real-life data is often noisy due to data collection errors, unusual user behavior, bot traffic, etc. Data preprocessing can partially eliminate this noise, enabling better modeling. E.g.~step (3) of the preprocessing described in section~\ref{ssec:flaw-data} and step (2) for Rees46. Preprocessing can also be used to adapt the dataset to the selected task and evaluation setup. E.g.~step (1) and (4) of our preprocessing. However, it is important to consider if preprocessing affects (a) how the outcome of the experiment can be interpreted; (b) whether the results are directly comparable with earlier work; (c) and if the experiment is still suitable to support the articulated claims. 

Modifying only the training set usually does not hurt the generality of claims about model performance, but direct comparison with earlier work and interpretation of the results might become non-trivial, since changes might affect algorithms differently. E.g.~shrinking the training window and using more recent data might benefit recommenders~\cite{tan2016improved} until the uplift from reduced concept drift is balanced out by the degradation from training on less data. Offline evaluation is already biased towards memorization type algorithms, e.g.~neighbor methods. This bias is stronger on smaller training windows due to more static user behavior and because generalization requires more data. Thus, reduced training window is likely to have more severe effect on more complex algorithms with better generalization capabilities. Figure~\ref{fig:gru_vs_sknn} demonstrates this by measuring the performance of the model-based GRU4Rec and the neighbor-based V-SkNN~\cite{ludewig2018evaluation} algorithms using training windows of varying sizes\footnote{Hyperparameters are optimized separately for each training window size and algorithm.}. While the performance of both models decrease as the training window shortens, the model-based method is affected more severely: GRU4Rec outperforms V-SkNN by $5.7\%$ ($8.9\%$) in recall@5 (recall@20) when both are trained on the full training set, but performs $29.6\%$ ($25.1\%$) worse when trained on the last 14 days. Unfortunately, connections between properties of training sets and their effect on the correlation between offline and online performance is not known, thus there is no ``right way'' to set the window size. However, biases like this should be considered when designing an evaluation setup.

Modifying the test set creates an entirely new evaluation setup, and heavy changes might result in less general performance claims. E.g.~evaluating on test users with more than 200 events is informative on the performance on established users only. Real-life recommenders should work for every user, but they can contain multiple algorithms. Therefore, evaluating on certain subsets is not a flaw, if the validity of claims is clearly communicated. Making only the most necessary preprocessing steps is advisable, because the stronger the filtering, the less general the claims can be. Unfortunately, this advice is often ignored~\cite{tang2018personalized,huang2018improving,he2017neural,he2016fast,wang2019neural}. 

\subsection{Information leaking through time}\label{ssec:flaw-split}

\begin{figure*}[!ht]
    \centering
    \begin{subfigure}[b]{.45\textwidth}
        \includegraphics[width=\textwidth]{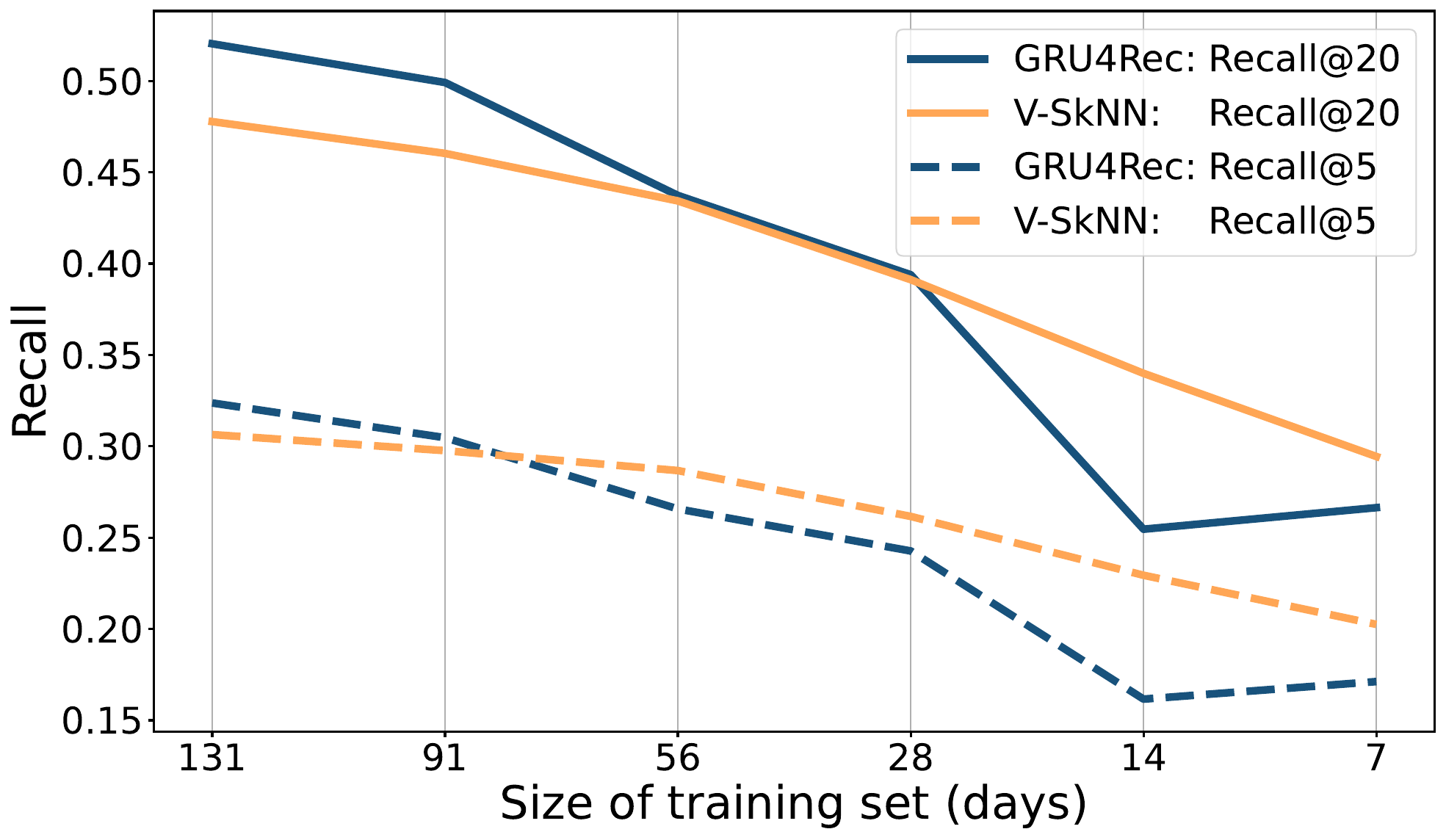}
        \Description[Accuracy measurements over different training windows]{Recall at 5 and 20 values measured for memorization and generalization type algorithms over different training windows.}
        \caption{The effect of using only recent data on the recommendation accuracy of model and neighbor based methods}
        \label{fig:gru_vs_sknn}
    \end{subfigure}
    \begin{subfigure}[b]{.45\textwidth}
        \includegraphics[width=\textwidth]{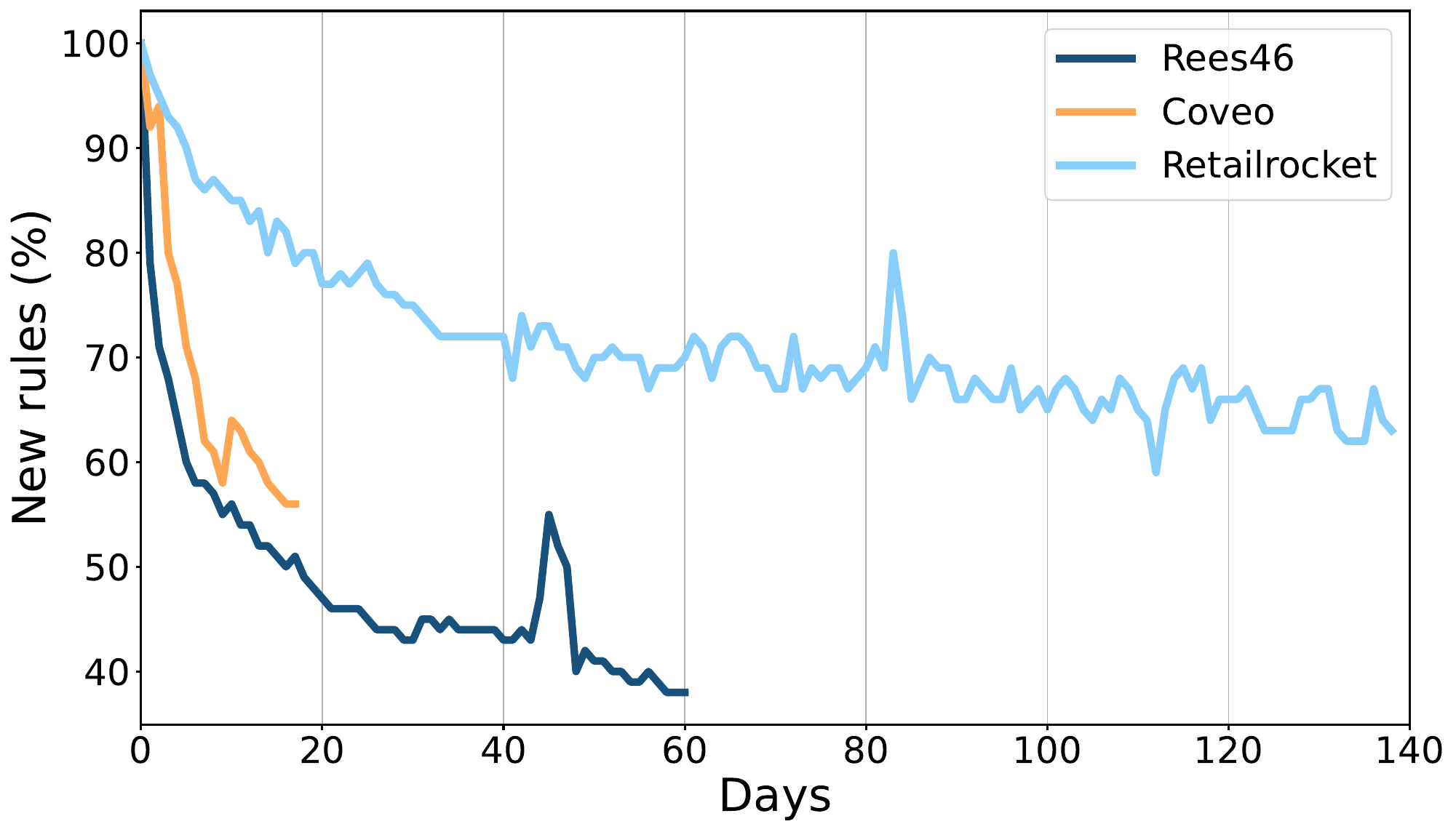}
        \Description[New rule proportion stops decreasing]{After a quick drop, the proportion of new rules stabilizes at a high level.}
        \caption{Proportion of $i\rightarrow j$ item transitions observed first on day $N$ to the number of unique sequences of the same day}
        \label{fig:new_rules}
    \end{subfigure}
    \caption{}
    \label{fig:combined_figure}
\end{figure*}

The next step in the design process is the train/test split. Proper evaluation requires eliminating access to any information that would not be available during inference. E.g.~training examples must not be used for evaluation. Recommenders should also not have access to future information during evaluation. However, some commonly used splitting strategies (e.g.~random split, leave-one-out) inherently enable information leaking through time.

Random splits are appropriate if the preference of users is considered to be long-term and mostly static. This is not the case due to changes in the item catalog, shifting user interest and external factors (e.g.~promotions). Figure~\ref{fig:new_rules} shows the proportion of $i\rightarrow j$ item transitions observed first on day $N$ and the number of unique sequences on the same day. While it declines, it stabilizes on fairly high values, indicating constantly changing user behavior. This concept-drift has been studied by the research community~\cite{gama2014survey,tsymbal2004problem}. Overlapping train/test splits lessen the need of modeling the concept drift, and provide an easier, unrealistic problem for which less generalization is needed. Therefore,  its effect is not uniform over different types of algorithms. 

\begin{figure*}[!h]
    \centering
    \begin{subfigure}[b]{.45\textwidth}
        \includegraphics[width=\textwidth]{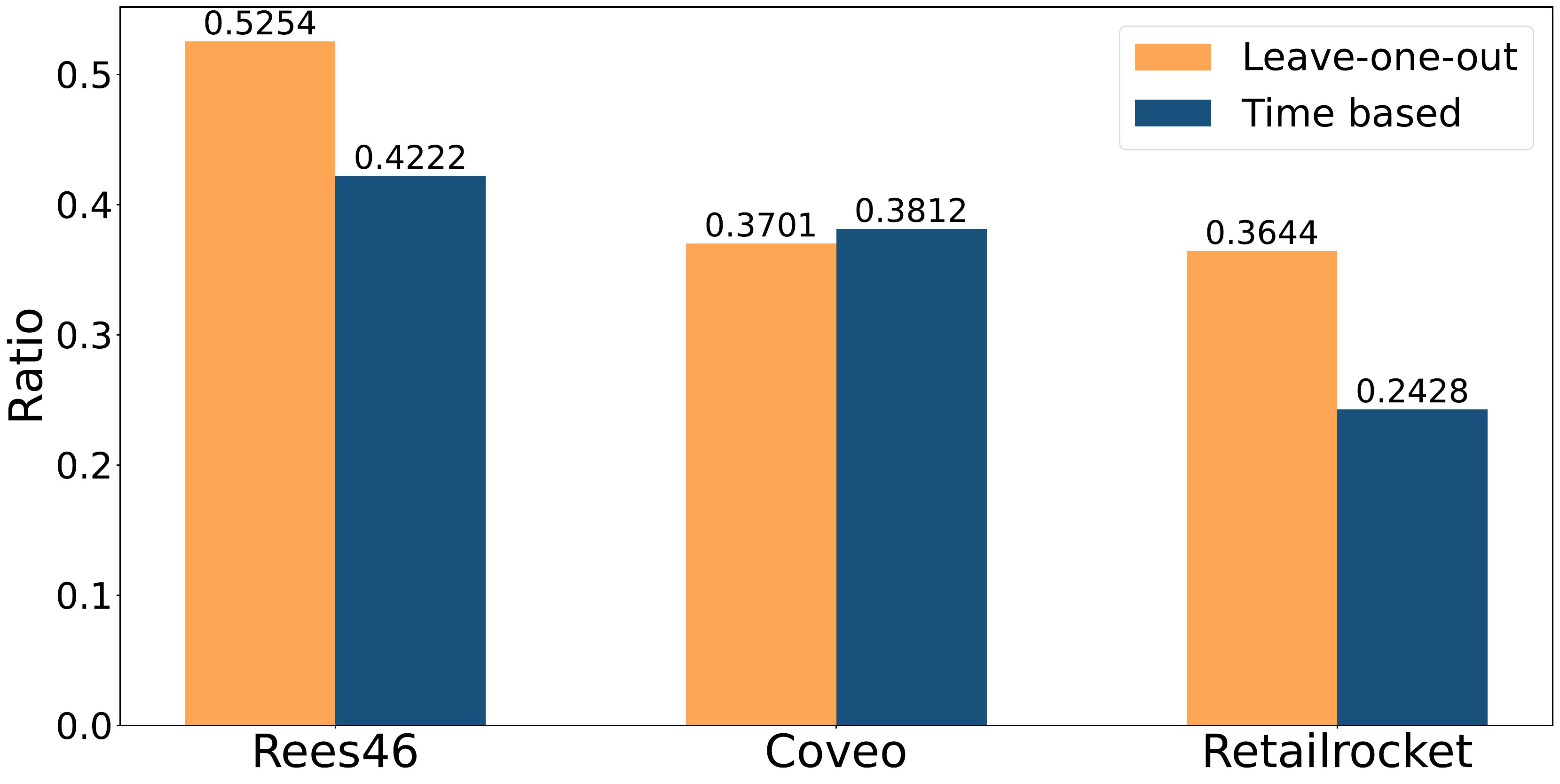}
        \Description[Bar plot of shared rules]{Bar plot of shared rules between training and test sets with leave-one-out having higher overlap}
        \caption{Leave-one-out and time based split}
        \label{fig:overlap_lo1_time}
    \end{subfigure}
    \begin{subfigure}[b]{.45\textwidth}
        \includegraphics[width=\textwidth]{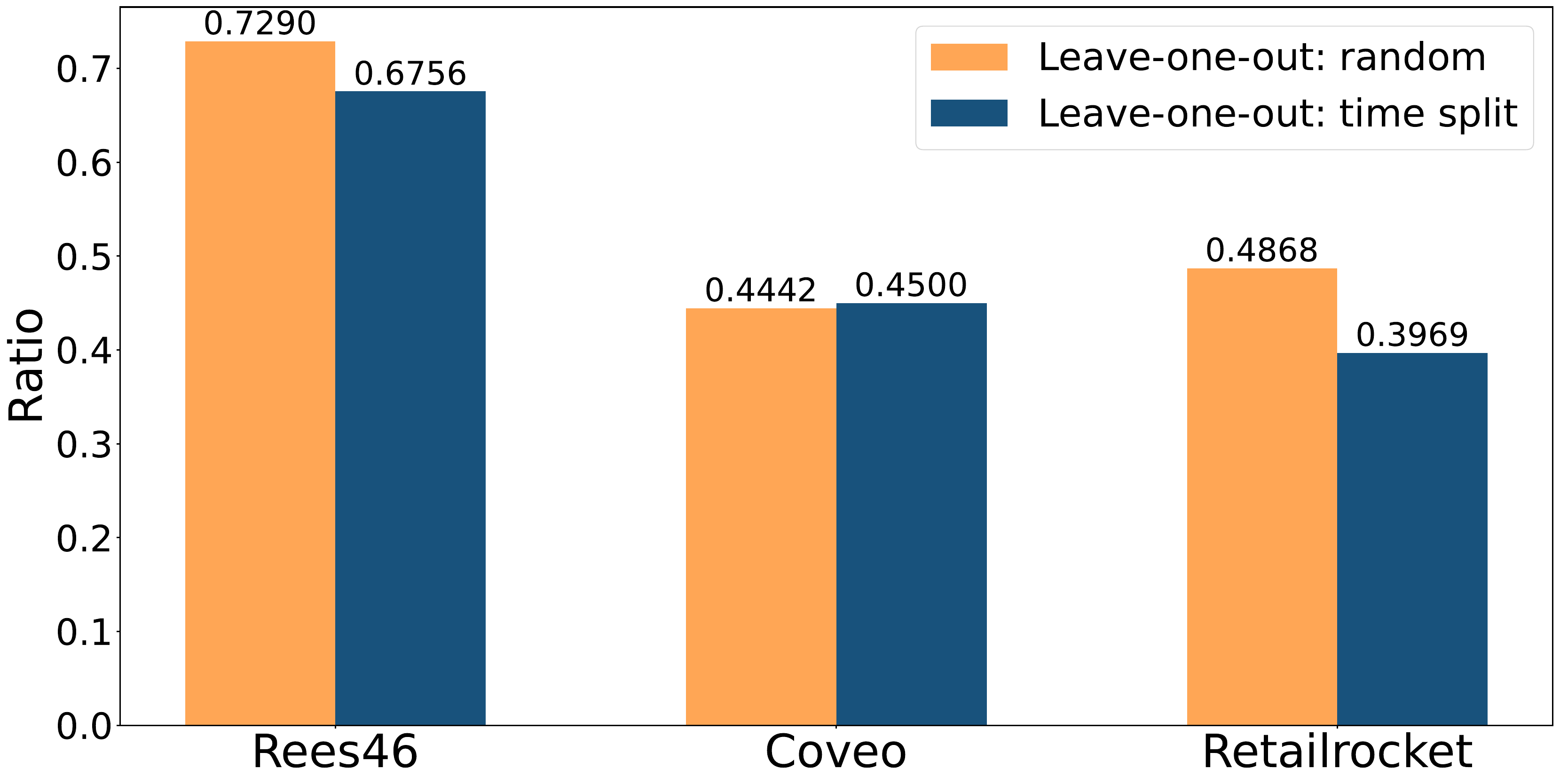}\Description[Bar plot of shared rules]{Bar plot of shared rules between training and test sets with random split having higher overlap}
        \caption{Leave-one-out on random vs. most recent sessions}
        \label{fig:overlap_random_vs_time_lo1}
    \end{subfigure}
    \caption{Proportion of the $i\rightarrow j$ test item transitions that are shared with the training set}
    \label{fig:dataset_split}
\end{figure*}

\emph{Leave-one-out} splitting is often used in the evaluation of sequential recommenders~\cite{kang2018self,sun2019bert4rec,huang2018improving,zhou2020s3,li2020time}. The last event of the training sequences are moved from the train set to the test set. Sequences end at different times, thus the two sets overlap in time. Figure~\ref{fig:dataset_split} shows the concept drift between train and test sets as the proportion of the $i\rightarrow j$ test item transitions that are shared with the train set\footnote{Note that the exact value also depends on the proportion of the training and test sets, therefore the size of the test sets was matched for both comparisons.} for leave-one-out and time based splits (\ref{fig:overlap_lo1_time}); and two variants of the leave-one-out strategy (\ref{fig:overlap_random_vs_time_lo1}) using the most recent and random sequences. Concept drift is significantly smaller for non-overlapping splits of Rees46 and RetailRocket. Coveo has similar proportions for both splits due to consisting of only 18 days of data, which is too short for concept drift to be too prevalent. This clearly indicates that by not requiring strict separation by time, the evaluation setup becomes compromised by information leaking through time, and thus the algorithms are evaluated on a somewhat easier problem than what they would face in a production system.

\subsection{Negative sampling during testing}\label{ssec:flaw-sampling}

Negative item sampling is a common and severe flaw of the final step, testing. Recently,~\cite{krichene2020sampled} discussed how sampling introduces bias to measurements and its potential impact on the comparison of algorithms, while~\cite{canamares2020target,dallmann2021case} demonstrated that it can change the performance based ordering of models. Thus, results achieved with sampling are not reliable. Besides confirming the result of earlier work, we demonstrate how the bias introduced by negative sampling can change the ordering of models, demonstrate its impact, as well as discuss alternatives and whether sampling is needed at all.

This flaw is rooted in the transition from error metrics to IR metrics from a decade ago that brought along a significant increase in the amount of compute needed for evaluation. Opposed to rating prediction that needs only a few score computations per test case, ranking requires scoring all items for every test case, since test items are treated as relevant ones that need to be ranked against other irrelevant items. Different alternatives~\cite{bellogin2011precision} were proposed including ranking over all items and ranking the target(s) against a number of sampled negative items. Ranking over all items is how real-life recommenders work, but nevertheless, negative sampling also became a widespread practice, utilized in well cited papers of prestigious conferences (e.g.~\cite{kang2018self,sun2019bert4rec,huang2018improving,zhou2020s3,he2017neural,cai2022aspect,rashed2022context}).

\begin{figure*}[!h]
    \centering
    \includegraphics[width=\textwidth]{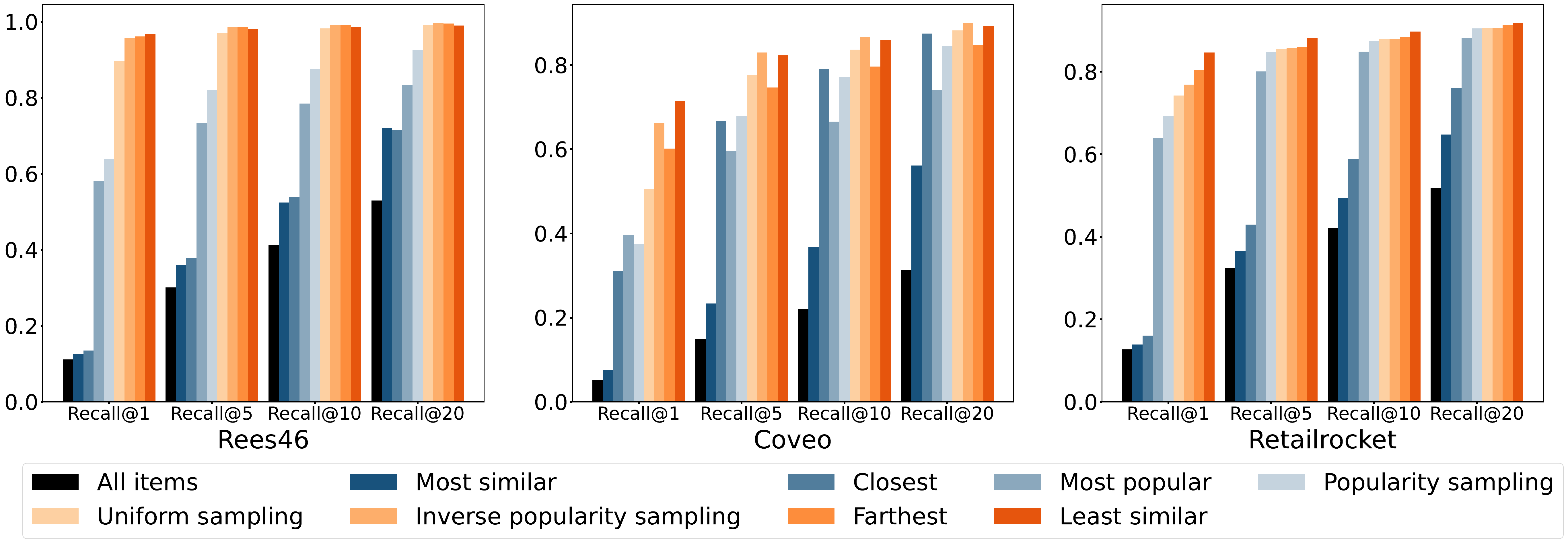}
    \Description[Bar plot of recall measurements]{Bar plot of recall measurements showing that sampling strategies result in overestimating accuracy metrics.}
    \caption{Comparison of the strength of various negative samples of 100 items and no sampling.}
    \label{fig:sampling_strategies}
\end{figure*}

Metrics focus on the top few items, thus negative items should be highly ranked for sampled and full ranking results to be similar. Uniform sampling is not able to provide strong samples if the size of the item catalog is significantly larger than the number of samples. The probability of a target item ranked $R$ in the full ranking of $N$ items making it into the recommendation list of $C$ items ($C<R$) when ranked against $S$ negative samples is $\sum_{i=0}^{C-1}{\binom{R-1}{i}\binom{N-R}{S-i}/\binom{N-1}{S}}$. If the target item is ranked $1,490^{\textrm{th}}$ among $10,000$ items or $14,878^{\textrm{th}}$ among $100,000$, it still has more than $90\%$ chance to make it into the top $20$ with $100$ uniform samples. Weak negative samples are easily distinguishable from the target and thus their use results in the severe overestimation of offline accuracy metrics. Using stronger samples provides a more accurate estimation of the accuracy of the non sampled setup. The strength of negative samples -- i.e.~how hard it is to distinguish them from the target -- is depicted on Figure~\ref{fig:sampling_strategies} for 8 sampling strategies:
\begin{itemize}
    \item \textbf{Sampling strategies:} Sampling probability is either uniform or proportional to the items' support. Popular items are considered to be stronger samples, since recommenders tend to rank them higher. 
    \item \textbf{Most popular items:} The target items are ranked against the $100$ most popular items.
    \item \textbf{Similar/close items:} Soft upper bound. Requires at least as much compute as full ranking. Items are selected based on the (cosine) similarity or closeness of their embeddings to that of the target item.
    \item \textbf{Weak baselines:} Soft lower bound. Sampling with probability proportional to reciprocal item support, and items with least similar or farthest embeddings to the target item's embedding. 
\end{itemize}
Figure~\ref{fig:sampling_strategies} indicates that even the computationally expensive most similar and closest approaches can not produce strong item sets consistently. Popularity sampling and most popular items are better than uniform sampling, but these approaches are still not able to estimate the true rank of the target. This can falsify the results, because (1) the difference between the performance of two algorithms can vanish due to the lack of challenging negative items. (2) The relative performance at recommendation list length $M$ shifts to length $N(\ll M)$ because sampling makes it easier to push the target item up on the list. Since the relative performance of two models might depend on the length of the recommendation list, this can also change their ordering for relevant list lengths. 

\begin{figure*}[!h]
    \centering
    \begin{subfigure}[b]{.32\textwidth}
        \includegraphics[width=\textwidth]{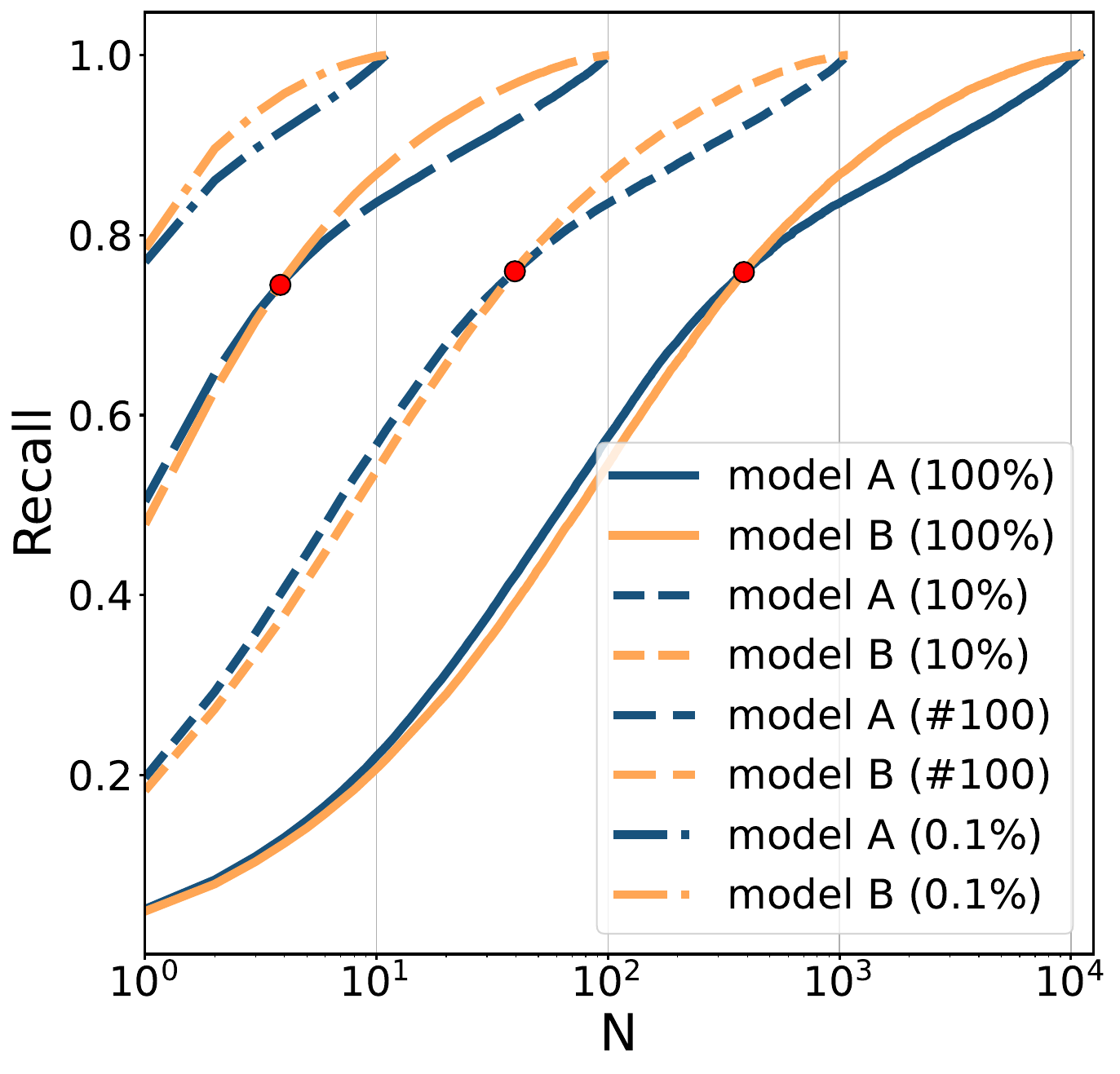}
        \Description[Recall values at different list length]{Recall values at different list length. Algorithm ranking changes at smaller list length as sample size decreases.}
        \caption{Coveo -- Recall@N}
        \label{fig:coveo_recall}
    \end{subfigure}
    \begin{subfigure}[b]{.32\textwidth}
        \includegraphics[width=\textwidth]{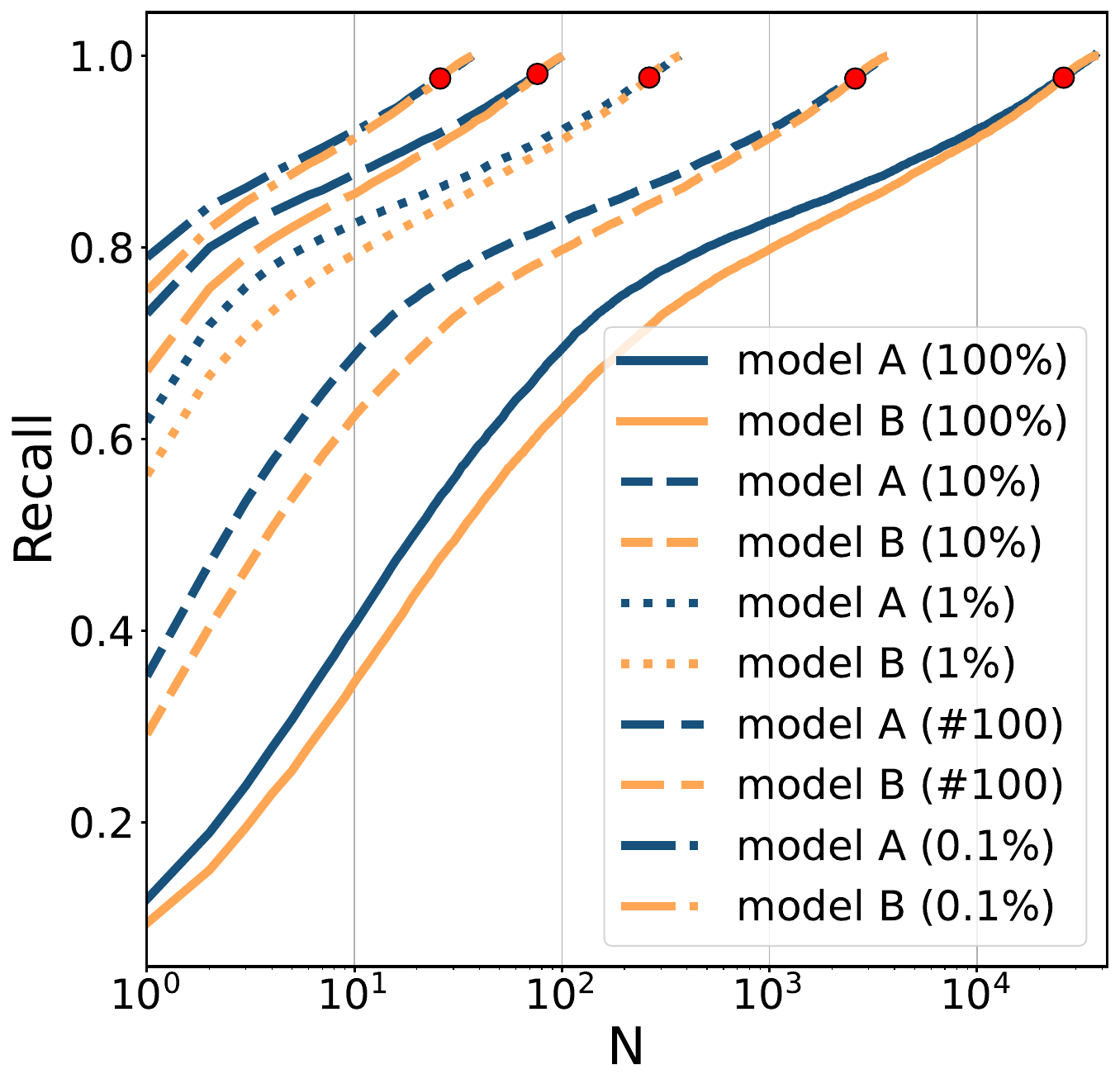}
        \Description[Recall values at different list length]{Recall values at different list length. Algorithm ranking changes at smaller list length as sample size decreases.}
        \caption{Retailrocket -- Recall@N}
        \label{fig:retailrocket_recall}
    \end{subfigure}
    \begin{subfigure}[b]{.32\textwidth}
        \includegraphics[width=\textwidth]{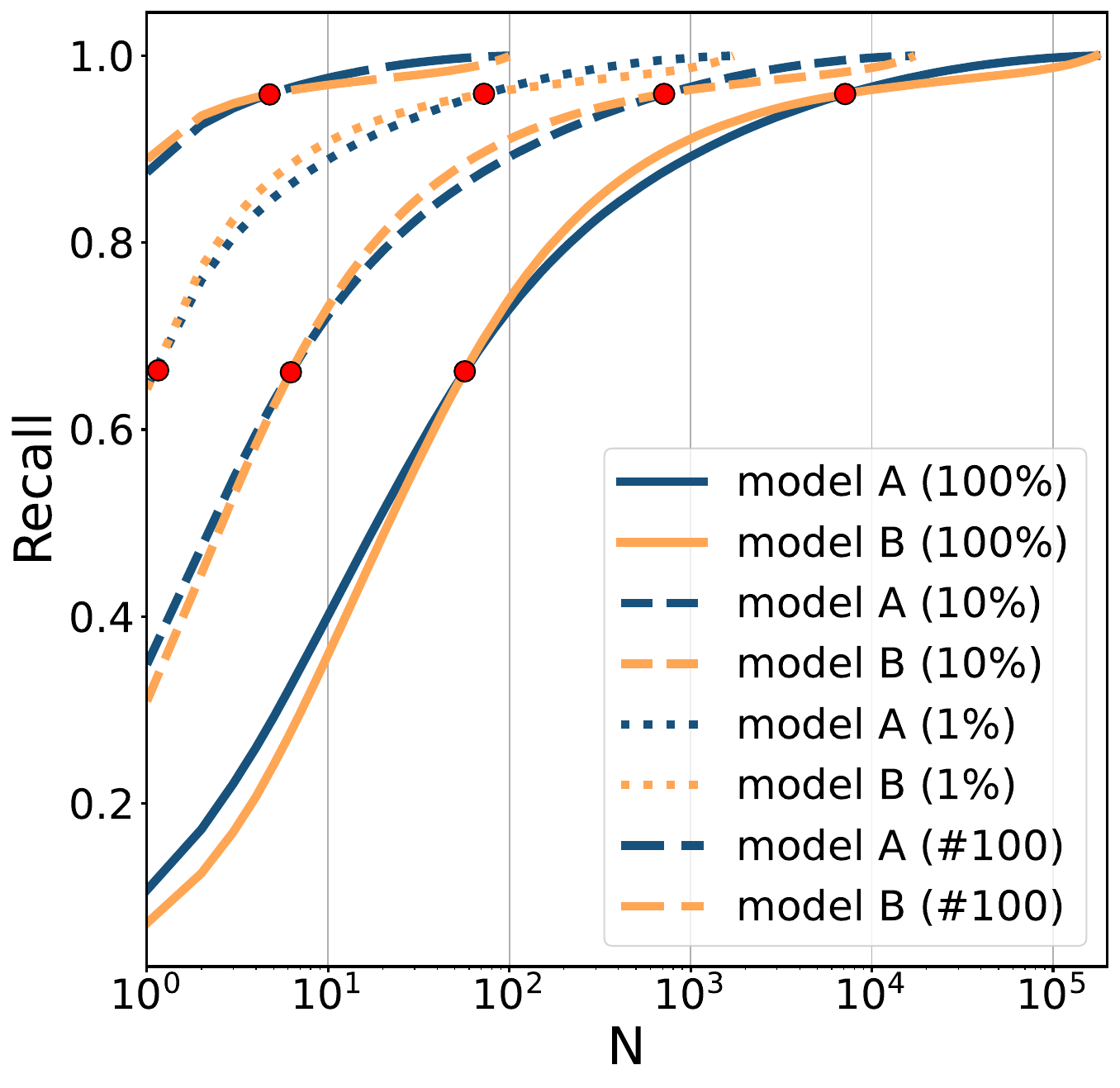}
        \Description[Recall values at different list length]{Recall values at different list length. Algorithm ranking changes at smaller list length as sample size decreases.}
        \caption{Rees46 -- Recall@N -- AB}
        \label{fig:rees46_recall_AB}
    \end{subfigure}
    \begin{subfigure}[b]{.32\textwidth}
        \includegraphics[width=\textwidth]{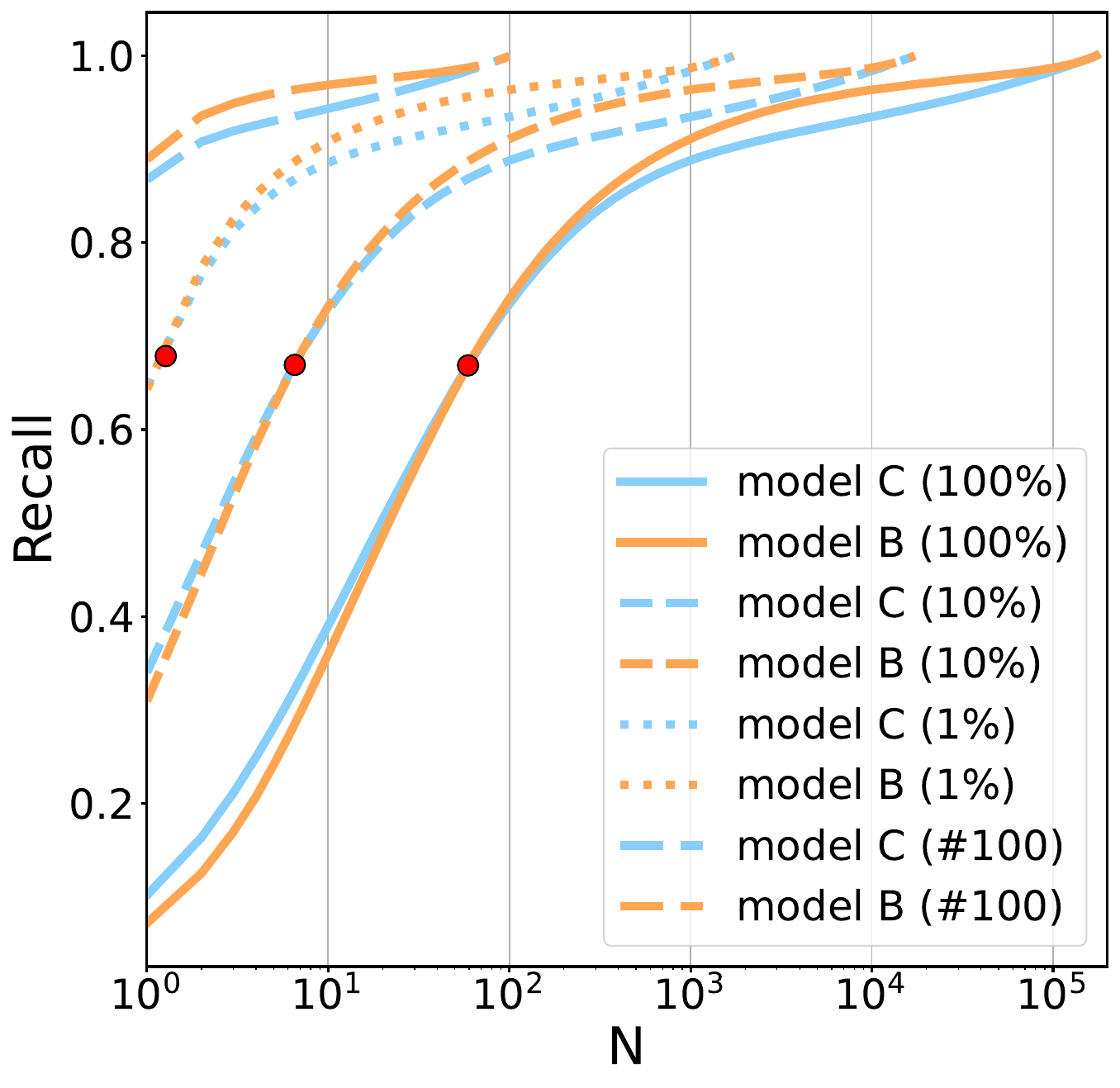}
        \Description[Recall values at different list length]{Recall values at different list length. Algorithm ranking changes at smaller list length as sample size decreases.}
        \caption{Rees46 -- Recall@N -- BC}
        \label{fig:rees46_recall_BC}
    \end{subfigure}
    \begin{subfigure}[b]{.32\textwidth}
        \includegraphics[width=\textwidth]{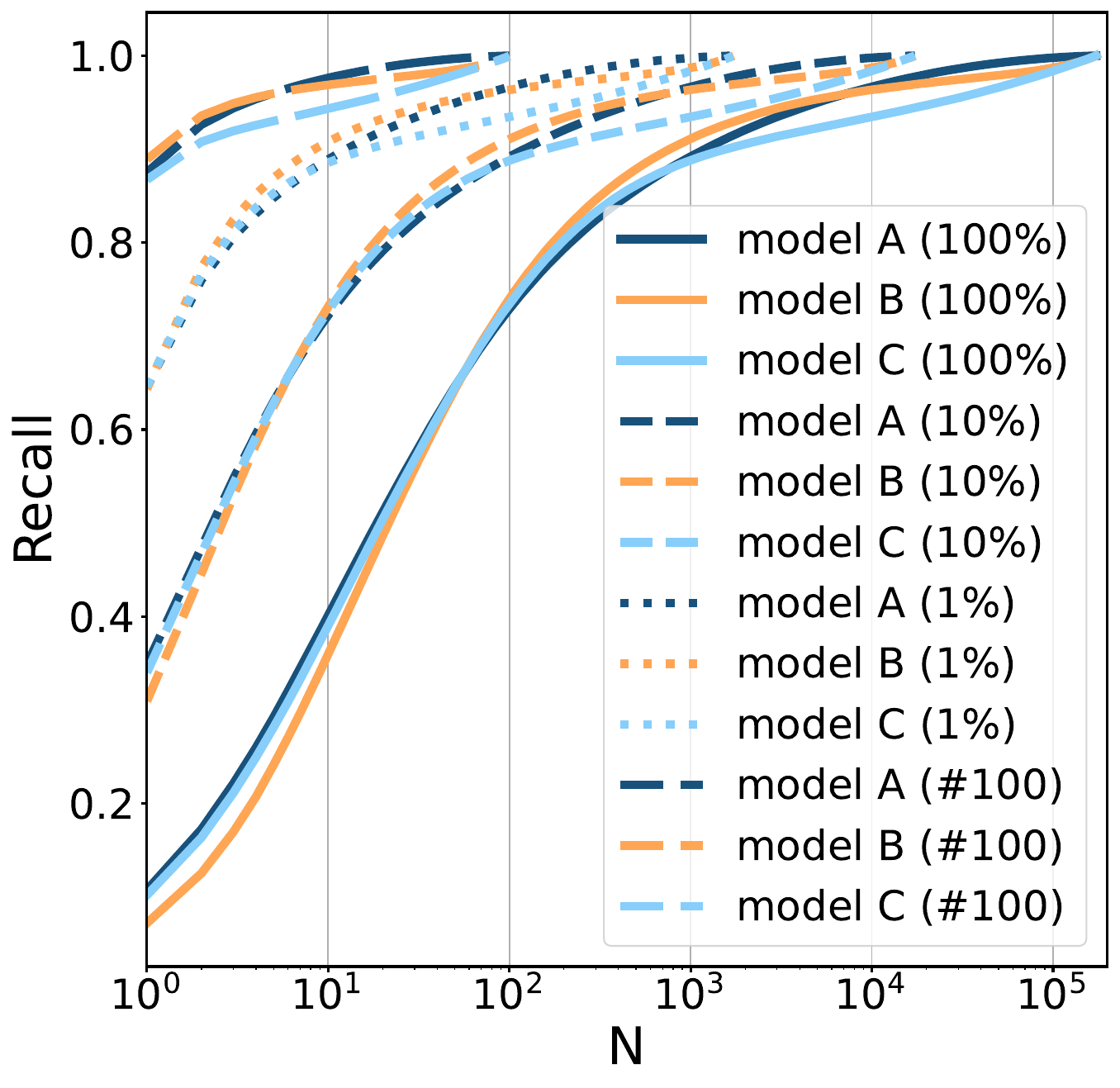}
        \Description[Recall values at different list length]{Recall values at different list length. Algorithm ranking changes at smaller list length as sample size decreases.}
        \caption{Rees46 -- Recall@N -- ALL}
        \label{fig:rees46_recall_ABC}
    \end{subfigure}
    \begin{subfigure}[b]{.32\textwidth}
        \includegraphics[width=\textwidth]{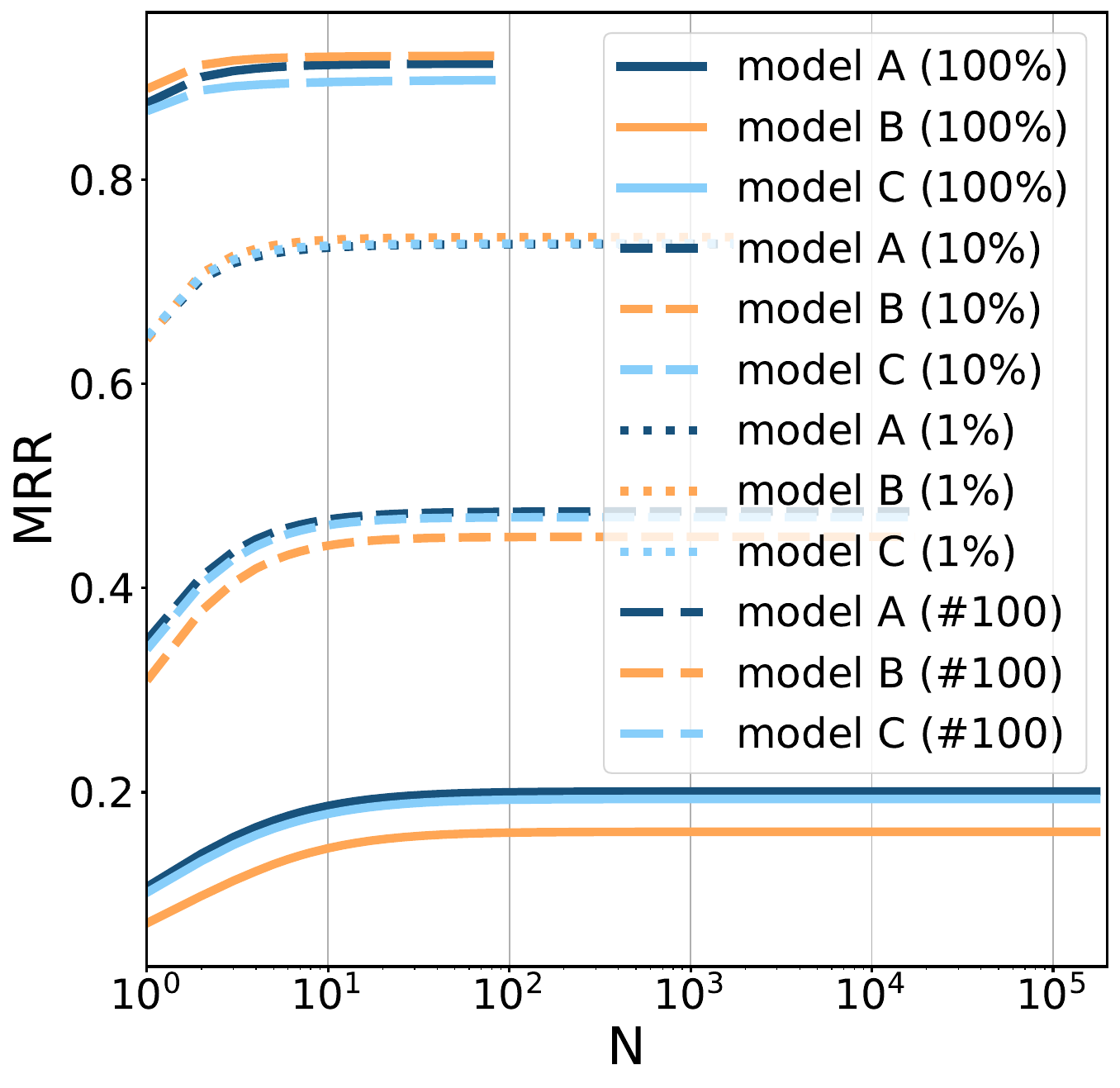}
        \Description[MRR values at different list length]{MRR values at different list length. Algorithm ranking changes as sample size decreases.}
        \caption{Rees46 -- MRR@N -- ALL}
        \label{fig:rees46_mrr}
    \end{subfigure}
    \caption{Accuracy as the function of recommendation list length, with and without sampling}
    \label{fig:metrics_at_n}
\end{figure*}

Figure~\ref{fig:metrics_at_n} demonstrates this point by showing recall@N as the function of $N$ for pairs of models for full ranking ($100\%$), and for randomly sampling $10\%$, $1\%$, $0.1\%$ of the item catalog or $100$ items. The ordering of models is not static over $N$, and it might change multiple times (marked by red dots). Decreasing sample size moves these intersections to the left, pushing the change in model ordering to lower $N$ values. Changes to the relative performance are drastic: model A on Coveo (\ref{fig:coveo_recall}) outperforms model B by $8.1\%$ in recall@20, but A underperforms B by $4.7\%$ when $100$ negative samples are used. While the ordering of model A and B does not change for Retailrocket (\ref{fig:retailrocket_recall}), $14.3\%$ uplift in recall@20 vanishes and drops to $1.5\%$. The biggest change can be observed between model A and B on Rees46 (\ref{fig:rees46_recall_AB}) where the uplift in recall@1 (and MRR@1) drops from $49.5\%$ to $-1.6\%$. Since MRR is even more sensitive to the top of the list, this affects relative MRR based performance for any $N$: the true order of models A, B and C on Rees46 (\ref{fig:rees46_mrr}) is $A>C\gg B$, but with $100$ samples it reads as $B>A\gg C$.

Computational complexity of evaluation via full ranking depends on the number of test recommendations and the size of the item catalog. Most public dataset are quite small, and even the bigger ones have item catalogs of easily manageable sizes. One test on Rees46 requires $887,369$ recommendation lists over $172,756$ items each that is executed in only $205.1$ seconds on an A30 GPU. Even before GPUs, evaluation on public datasets of the time could be executed in a few minutes using optimized code. If a dataset is really too large for quick experimentation, the sampling of test recommendations (e.g.~users) gives more representative results. Certain scoring models are not designed for full ranking. If one such model must be evaluated on ranking, generating candidate sets via another algorithm can solve the problem. However, this limits the generality of claims about its performance, because currently there are no universally accepted candidate set generators.

%% file: 3_related_work.tex
\section{Related work}\label{sec:related}

Offline evaluation of top-N recommenders has been discussed since top-N recommendation became dominant. Early works mainly focused on metrics~\cite{hurley2011novelty,vargas2011rank} and finding connections between offline and online results~\cite{10.1145/2959100.2959176}. The pace of the work has increased in recent years~\cite{10.1145/3523227.3547408}, simultaneously with the sharp increase of papers lacking satisfactory evaluation. 

Evaluation has many aspects from metrics~\cite{hurley2011novelty,vargas2011rank} to hyperparameter optimization~\cite{rendle2022revisiting}. \cite{ferrari2019we,ferrari2020methodological} suggests that one of the reasons behind the low reproducibility rate of recent papers is that evaluation setups are lifted from earlier work without their validity being checked. 

Our work focuses on this problem, i.e.~the design of evaluation setups and the flaws within. Of the four flaws discussed in this paper, only negative item sampling has been scrutinized, aside from \cite{tang2018personalized} briefly mentioning that the Amazon dataset has weak sequential signals. \cite{krichene2020sampled} discussed the background of how sampling introduces bias to measurements, and~\cite{canamares2020target,dallmann2021case} demonstrated that the relative performance of models can change when sampling is applied. Our experiments confirm their results and give additional explanation on why this change happens.

%% file: 4_conclusion.tex
\section{Discussion \& Conclusion}\label{sec:conclusion}

Offline evaluation is fundamentally imperfect, but it will likely remain the main approach for assessing performance in recommendation systems research. Online A/B tests are not just expensive and slow, but inherently not reproducible; and simulators are still in their early stages. Unfortunately, flaws are quite widespread in offline evaluation setups. In this paper, we went through the main steps of designing evaluation setups and pointed out four widespread flaws and demonstrated their effect through the example of sequential recommendations. We chose sequential recommendation for the demonstration, because it is one of the areas severely plagued by low quality evaluation. Evaluation flaws of some of the earlier sequential recommender papers -- e.g.~\cite{kang2018self} has all four we discussed -- were missed by both the authors and reviewers. Later works copied the setups -- with their flaws included -- without questioning whether they are correct. There is not a single best way for offline evaluation, but by pointing out these four flaws, we hope to see the quality of experimentation sections of research papers improving in the future.